%% file: paper2p.tex
\documentclass[a4paper]{article}
\usepackage{graphicx}
\usepackage{twocolpceurws}
\usepackage{multirow}

\title{Word Embeddings to Enhance Twitter Gang Member Profile Identification}

\author{
Sanjaya Wijeratne \\  sanjaya@knoesis.org
\and
Lakshika Balasuriya \\  lakshika@knoesis.org
\and
Derek Doran \\  derek@knoesis.org
\and
Amit Sheth \\  amit@knoesis.org
}

\institution{Kno.e.sis Center, Wright State University, Dayton, Ohio, USA.}

\begin{document}
\maketitle

\begin{abstract}
\textbf{Gang affiliates have joined the masses who use social media to share thoughts and actions publicly. Interestingly, they use this public medium to express recent illegal actions, to intimidate others, and to share outrageous images and statements. Agencies able to unearth these profiles may thus be able to anticipate, stop, or hasten the investigation of gang-related crimes. This paper investigates the use of word embeddings to help identify gang members on Twitter. Building on our previous work, we generate word embeddings that translate what Twitter users post in their profile descriptions, tweets, profile images, and linked YouTube content to a real vector format amenable for machine learning classification. Our experimental results show that pre-trained word embeddings can boost the accuracy of supervised learning algorithms trained over gang members' social media posts.}
\end{abstract}

\section{Introduction}
Street gangs are defined as ``a coalition of peers, united by mutual interests, with identifiable leadership and internal organization, who act collectively to conduct illegal activity and to control a territory, facility, or enterprise''~\cite{miller1992crime}. They promote criminal activities such as drug trafficking, assault, robbery, and threatening or intimidating a neighborhood~\cite{2013NationalGangReport}. Today, over 1.4 million people, belonging to more than 33,000 gangs, are active in the United States~\cite{2011NationalGangThreat}, of which 88\% identify themselves as being members of a street gang\footnote{The terms `gang' and `street gang' will henceforth be used interchangeably.}. They are also active users of social media~\cite{2011NationalGangThreat}; according to 2007 National Assessment Center's survey of gang members, 25\% of individuals in gangs use the Internet for at least 4 hours a week~\cite{2007assesment}. More recent studies report approximately 45\% of gang members participate in online offending activities such as threatening, harassing individuals, posting violent videos or attacking someone on the street for something they said online~\cite{decker2011leaving,doi:10.1080/07418825.2013.778326}. They confirm that gang members use social media to express themselves in ways similar to their offline behavior on the streets~\cite{patton13}.

Despite its public nature, gang members post on social media without fear of consequences because there are only few tools law enforcement can presently use to surveil social media~\cite{7165945}. For example, the New York City police department employs over 300 detectives to combat teen violence triggered by insults, dares, and threats exchanged on social media, and the Toronto police department teaches officers about the use of social media in investigations~\cite{police13}. From offline clues, the officers monitor just a selected set of social media accounts which are manually discovered and  related to a specific investigation. Thus, developing tools to identify gang member profiles on social media is an important step in the direction of using machine intelligence to fight crime.

To help agencies monitor gang activity on social media, our past work investigated how features from Twitter profiles, including profile text, profile images, tweet text, emjoi use, and their links to YouTube, may be used to reliably find gang member profiles~\cite{balasuriya2016twittergang}. The diverse set of features, chosen to combat the fact that gang members often use local terms and hashtags in their posts, offered encouraging results. In this paper, we report our experience in integrating deep learning into our gang member profile classifier. Specifically, we investigate the effect of translating the features into a vector space using word embeddings~\cite{NIPS2013_5021}. This idea is motivated by the recent success of word embeddings-based methods to learn syntactic and semantic structures automatically when provided with large datasets. A dataset of over 3,000 gang and non-gang member profiles that we previously curated is used to train the word embeddings. We show that pre-trained word embeddings improve the machine learning models and help us obtain an $F1$-score of $0.7835$ on gang member profiles (a 6.39\% improvement in $F1$-score compared to the baseline models which were not trained using word embeddings).

This paper is organized as follows. Section~\ref{sec:rr} discusses the related literature and frames how this work differs from other related works. Section \ref{sec:dc} discusses our approach based on word embeddings to identify gang member profiles. Section \ref{sec:eval} reports on the evaluation of the proposed approach and the evaluation results in detail. Section \ref{sec:con} concludes the work reported while discussing the future work planned.

\section{Related Work}\label{sec:rr}
Researchers have begun investigating the gang members' use of social media and have noticed the importance of identifying gang members' Twitter profiles a priori~\cite{patton13,7165945}. Before analyzing any textual context retrieved from their social media posts, knowing that a post has originated from a gang member could help systems to better understand the message conveyed by that post. Wijeratne {\em et al.} developed a framework to analyze what gang members post on social media~\cite{7165945}. Their framework could only extract social media posts from self identified gang members by searching for pre-identified gang names in a user's Twitter profile description. Patton {\em et al.} developed a method to collect tweets from a group of gang members operating in Detroit, MI~\cite{desmondupatton2015}. However, their approach required the gang members' Twitter profile names to be known beforehand, and data collection was localized to a single city in the country. These studies investigated a small set of manually curated gang member profiles, often from a small geographic area that may bias their findings.

In our previous work~\cite{balasuriya2016twittergang}, we curated what may be the largest set of gang member profiles to study how gang member Twitter profiles can be automatically identified based on the content they share online. A data collection process involving location neutral keywords used by gang members, with an expanded search of their retweet, friends and follower networks, led to identifying 400 authentic gang member profiles on Twitter. Our study discovered that the text in their tweets and profile descriptions, their emoji use, their profile images, and music interests embodied by links to YouTube music videos, can help a classifier distinguish between gang and non-gang member profiles. While a very promising $F1$ measure with low false positive rate was achieved, we hypothesize that the diverse kinds and the multitude of features employed (e.g. unigrams of tweet text) could be amenable to an improved representation for classification. We thus explore the possibility of mapping these features into a considerably smaller feature space through the use of word embeddings.

Previous research has shown word embeddings-based methods can significantly improve short text classification~\cite{Wang2016806,7259377}. For example, Lilleberget {\em et al.} showed that word embeddings weighted by $tf$-$idf$ outperforms other variants of word embedding models discussed in~\cite{7259377}, after training word embedding models on over 18,000 newsgroup posts. Wang {\em et al.} showed that short text categorization can be improved by word embeddings with the help of a neural network model that feeds semantic cliques learned over word embeddings in to a convolutions neural network~\cite{Wang2016806}. We believe our corpus of gang and non-gang member tweets, with nearly 64.6 million word tokens, could act as a rich resource to train word embeddings for distinguishing gang and non-gang member Twitter users. Our investigation differs from other word embeddings-based text classification systems such as~\cite{Wang2016806,7259377} due to the fact that we use multiple feature types including emojis in tweets and image tags extracted from Twitter profile and cover images in our classification task.

\section{Word Embeddings} \label{sec:dc}
A word embedding model is a neural network that learns rich representations of words in a text corpus. It takes data from a large, $n$-dimensional `word space' (where $n$ is the number of unique words in a corpus) and learns a transformation of the data into a lower $k$-dimensional space of real numbers. This transformation  is developed in a way that similarities between the $k$-dimensional vector representation of two words reflects semantic relationships among the words themselves. These semantics are not captured by typical bag-of-words or $n$-gram models for classification tasks on text data~\cite{mikolov-yih-zweig:2013:NAACL-HLT,NIPS2013_5021}.

Word embeddings have led to the state-of-the-art results in many sequential learning tasks~\cite{lecun2015deep}. In fact, word embedding learning is an important step for many statistical language modeling tasks in text processing systems. Bengio {\em et al.} were the first ones to introduce the idea of learning a distributed representation for words over a text corpus~\cite{Bengio:2003:NPL:944919.944966}. They learned representations for each word in the text corpus using a neural network model that modeled the joint probability function of word sequences in terms of the feature vectors of the words in the sequence. Mikolov {\em et al.} showed that simple algebraic operations can be performed on word embeddings learned over a text corpus, which leads to findings such as the word embedding vector of the word ``King'' $-$ the word embedding vectors of ``Man'' $+$ ``Woman'' would results in a word embedding vector that is closest to the word embedding vector of the word ``Queen''~\cite{mikolov-yih-zweig:2013:NAACL-HLT}. Recent successes in using word embeddings to improve text classification for short text~\cite{Wang2016806,7259377}, encouraged us to explore how they can be used to improve gang and non-gang member Twitter profile classification.

Word embeddings can be performed under different neural network architectures; two popular ones are the Continuous Bag-of-Words (CBOW) and Continuous Skip-gram (Skip-gram) models~\cite{DBLP:journals/corr/abs-1301-3781}. The CBOW model learns a neural network such that given a set of context words surrounding a target word, it predict a target word. The Skip-gram model differs by predicting context words given a target word and by capturing the ordering of word occurrences. Recent improvements to Skip-gram model make it better able to handle less frequent words, especially when negative sampling is used~\cite{NIPS2013_5021}.

\subsection{Features considered} \label{features}
Gang member tweets and profile descriptions tend to have few textual indicators that demonstrate their gang affiliations or their tweets/profile text may carry acronyms which can only be deciphered by others involved in gang culture~\cite{balasuriya2016twittergang}. These gang-related terms are often local to gangs operating in neighborhoods and change rapidly when they form new gangs. Consequently, building a database of keywords, phrases, and other identifiers to find gang members nationally is not feasible. Instead, we use heterogeneous sets of features derived not only from profile and tweet text but also from the emoji usage, profile images, and links to YouTube videos reflecting their music preferences and affinity. In this section, we briefly discuss the feature types and their broad differences in gang and non-gang member profiles. An in-depth explanation of these feature selection can be found in~\cite{balasuriya2016twittergang}.

\begin{enumerate} 

\item  {\bf Tweet text:} In our previous work, we observed that gang members use curse words nearly five times more than the average curse words use on Twitter~\cite{balasuriya2016twittergang}. Further, we noticed that gang members mainly use Twitter to discuss drugs and money using terms such as {\em smoke, high, hit, money, got,} and {\em need} while non-gang members mainly discuss their feelings using terms such as {\em new, like, love, know, want,} and {\em  look}.

\item  {\bf Twitter profile description:} We found gang member profile descriptions to be rife with curse words ({\em nigga, fuck,} and {\em  shit}) while non-gang members use words related to their feelings or interests ({\em love, life, music,} and {\em  book}). We noticed that gang members use their profile descriptions as a space to grieve for their fallen or incarcerated gang members as about $12\%$ of gang member Twitter profiles used terms such as {\em rip} and {\em free}. 

\item {\bf Emoji features:} We found that the fuel pump emoji was the most frequently used emoji by gang members, which is often used in the context of selling or consuming marijuana. The pistol emoji was the second most frequently used emoji, which is often used with the police cop emoji in an `emoji chain' to express their hatred towards law enforcement officers. The money bag emoji, money with wings emoji, unlock emoji, and a variety of the angry face emoji such as the devil face emoji and imp emoji were also common in gang members' but not in non-gang members' tweets. 

\item {\bf Twitter profile and cover images:} We noticed that gang members often pose holding or pointing weapons, seen in a group fashion which displays a gangster culture, show off graffiti, hand signs, tattoos, and bulk cash in their profile and cover images. We used Clarifai web service\footnote{http://www.clarifai.com/} to tag the profile and cover images of the Twitter users in our dataset and used the image tags returned by Clarifai API to train word embeddings. Tags such as {\em trigger, bullet,} and {\em worship} were unique for gang member profiles while non-gang member images had unique tags such as {\em beach, seashore, dawn, wildlife, sand,} and {\em pet}.

\item {\bf YouTube videos:} We found that 51.25\% of the gang members in our dataset have a tweet that links to a YouTube video. Further, we found that 76.58\% of the shared links are related to hip-hop music, gangster rap, and the culture that surrounds this music genre~\cite{balasuriya2016twittergang}. Moreover, we found that eight YouTube links are shared on average by a gang member. The top 5 terms used in YouTube videos shared by  gang members were {\em shit, like, nigga, fuck,} and {\em lil} while {\em like, love, peopl, song,} and {\em get} were the top 5 terms in non-gang members' video data.

\end{enumerate}

\subsection{Classification approach}\label{sec:approach}
\begin{figure}
\centering
\includegraphics[width=1.0\linewidth]{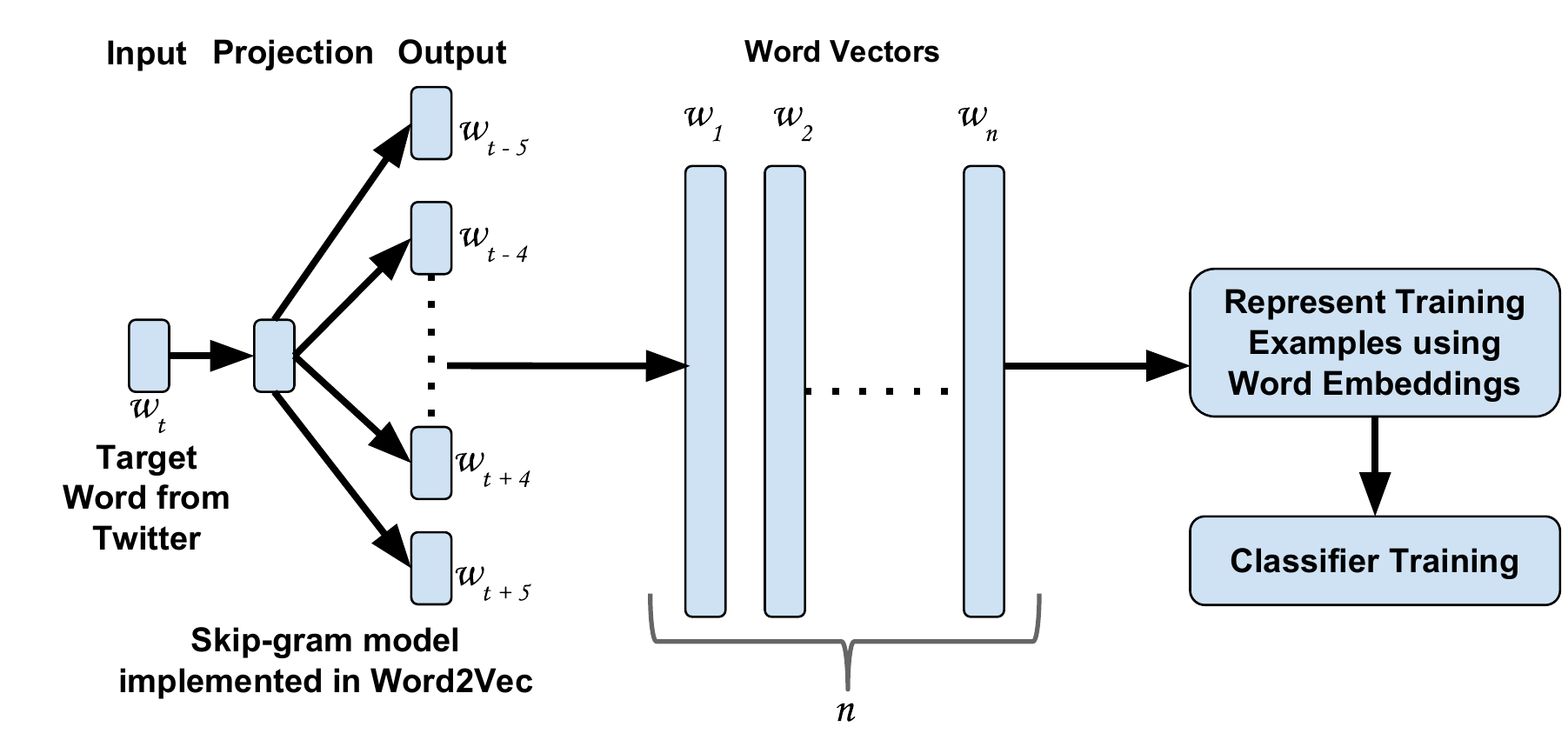} 
\caption{Classifier Training with Word Embeddings.}
\label{fig:approach}
\end{figure}

Figure~\ref{fig:approach} gives an overview of the steps to learn word embeddings and to integrate them into a classifier. We first convert any non-textual features such as emoji and profile images into textual features. We use Emoji for Python\footnote{https://pypi.python.org/pypi/emoji/} and Clarifai services, respectively, to convert emoji and images into text. Prior to training the word embeddings, we remove all the seed words used to find gang member profiles and stopwords, and perform stemming across all tweets and profile descriptions. We then feed all the training data (word $w_t$ in Figure~\ref{fig:approach}) we collected from our Twitter dataset to Word2Vec tool and train it using a Skip-gram model with negative sampling. When training the Skip-gram model, we set the negative sampling rate to 10 sample words, which seems to work well with medium-sized datasets~\cite{NIPS2013_5021}. We set the context word window to be 5, so that it will consider 5 words to left and right of the target word (words $w_{t-5}$ to $w_{t+5}$ in Figure~\ref{fig:approach}). This setting is suitable for sentences where average sentence length is less than 11 words, as is the case in tweets~\cite{HuTK13}. We ignore the words that occur less than 5 times in our training corpus.

We investigated how well the local language has been captured by the word embedding models we trained. We used the `most similar' functionality offered by Word2Vec tool to understand what the model has learned about few gang-related slang terms which are specific to Chicago area. For example, we analyzed the ten most similar words learned by the word embedding model for the term \texttt{BDK} (Black Desciples Killers). We noticed that out of the 10 most similar words, five were names of local Chicago gangs, which are rivals of the Black Disciples Gang, two were different syntactic variations of \texttt{BDK} ({\em bdkk, bdkkk}) and the other three were different syntactic variations of \texttt{GDK} ({\em gdk, gdkk, gdkkk}). \texttt{GDK} is a local gang slang for `Gangster Disciples Killer' which is used by rivals of Gangster Disciples gang to show their hatred towards them. We found similar results for the term \texttt{GDK}. Out of the ten most similar words, six were showing hatred towards six different Gangster Disciples gangs that operate in Chicago area. We believe that those who used the term \texttt{GDK} to show their hatred towards Gangster Disciples gangs might be also having rivalry with the six gangs we found. 

We obtain word vectors of size 300 from the learned word embeddings. To represent a Twitter profile, we retrieve word vectors for all the words that appear in a particular profile including the words appear in tweets, profile description, words extracted from emoji, cover and profile images converted to textual formats, and words extracted from YouTube video comments and descriptions for all YouTube videos shared in the user's timeline. Those word vectors are combined  to compute the final feature vector for the Twitter profile. To combine the word vectors, we consider five different methods. Letting the size of a word vector be $k = 300$, for a Twitter profile $p$ with $n$ unique words and the vector of the $i^{th}$ word in $p$ denoted by $w_{ip}$, we compute the feature vector for the Twitter profile $V_{p}$
by:

\begin{enumerate} 
  
  \item {\bf Sum of word embeddings} $V_{p_{sum}}$. This is the sum the word embedding vectors obtained for all words in a Twitter profile: 
  \[ V_{p_{sum}} = \sum_{i=0}^n w_{ip} \]

 \item {\bf Mean of word embeddings} $V_{p_{avg}}$. This is the mean of the word embedding vectors of all words found in a Twitter profile:
   \[ V_{p_{avg}} = 1/n\sum_{i=0}^n w_{ip} \]
 
  \item {\bf Sum of word embeddings weighted by term frequency} $V_{p_{sum(count)}}$. This is each word embedding vector multiplied by the word's frequency for the Twitter profile:
 \[ V_{p_{sum(count)}} = \sum_{i=0}^n w_{ip}.c_{ip} \] where $c_{ip}$ is the term frequency for the $i^{th}$ word in profile $p$. 
 
 \item {\bf Sum of word embeddings weighted by} $tf$-$idf$ $V_{p_{sum(tf-idf)}}$. This is each word vector  multiplied by the word's $tf$-$idf$ for the Twitter profile:
  \[ V_{p_{sum(tf-idf)}} = \sum_{i=0}^n w_{ip}.t_{ip} \] where $t_{ip}$ is the $tf$-$idf$ value for the $i^{th}$ word in profile $p$. 
 
  \item {\bf Mean of word embeddings weighted by term frequency} $V_{p_{avg(sum(count))}}$. This
  is the mean of the word embedding vectors weighted by term frequency:
   \[ V_{p_{avg(sum(count))}} = 1/n\sum_{i=0}^n w_{ip}.c_{ip} \]

 \end{enumerate}

\section{Evaluation} \label{sec:eval}
We evaluate the performance of using word embeddings to discover gang member profiles on Twitter. We first discuss the dataset, learning algorithms and baseline comparison models used in the experiments. Then we discuss the 10-fold cross validation experiments and the evaluation matrices used. Finally we present the results of the experiments.

\begin{table}[]
\label{datasetstats}
\begin{tabular}{|l|c|c|c|}
\hline
\multicolumn{1}{|c|}{\begin{tabular}[c]{@{}c@{}}\# of \\ Words in\end{tabular}} & \begin{tabular}[c]{@{}c@{}}Gang \\ Members\end{tabular} & \begin{tabular}[c]{@{}c@{}}Non-gang \\ Members\end{tabular} & \textbf{Total}      \\ \hline
Tweets                               & 3,825,092                                               & 45,213,027                                                  & \textbf{49,038,119} \\ \hline
Profiles                             & 3,348                                                   & 21,182                                                      & \textbf{24,530}     \\ \hline
Emoji                               & 732,712                                                 & 3,685,669                                                   & \textbf{4,418,381}  \\ \hline
Videos                       & 554,857                                                 & 10,459,235                                                  & \textbf{11,014,092} \\ \hline
Images                           & 10,162                                                  & 73,252                                                      & \textbf{83,414}     \\ \hline
\multicolumn{1}{|c|}{\textbf{Total}} & \textbf{5,126,171}                                      & \textbf{59,452,365}                                         & \textbf{64,578,536} \\ \hline
\end{tabular}
\centering
\caption{Dataset Statistics.}
\end{table}

\subsection{Evaluation setup}
We consider a dataset of curated gang and non-gang members' Twitter profiles collected from our previous work~\cite{balasuriya2016twittergang}. It was developed by querying the Followerwonk Web service API\footnote{https://moz.com/followerwonk/bio} with location-neutral seed words known to be used by gang members across the U.S. in their Twitter profiles. The dataset was further expanded by examining the friends, follower, and retweet networks of the gang member profiles found by searching for seed words. Specific details about our data curation procedure are discussed in~\cite{balasuriya2016twittergang}. Ultimately, this dataset consists of 400 gang member profiles and 2,865 non-gang member profiles. For each user profile, we collected up to most recent 3,200 tweets from their Twitter timelines, profile description text, profile and cover images, and the comments and video descriptions for every YouTube video shared by them. Table 1 provides statistics about the number of words found in each type of feature in the dataset. It includes a total of 821,412 tweets from gang members and 7,238,758 tweets from non-gang members.

To build the classifiers we used three different learning algorithms, namely Logistic Regression (LR), Random Forest (RF), and Support Vector Machines (SVM). We used version 0.17.1 of scikit-learn\footnote{http://scikit-learn.org/stable/index.html} machine learning library for Python to implement the classifiers. An open source tool of Python, Gensim~\cite{rehurek_lrec} was used to generate the word embeddings. We compare our results with the two best performing systems reported in~\cite{balasuriya2016twittergang} which are the two state-of-the-art models for identifying gang members in Twitter. Both baseline models are built from a random forest classifier trained over term frequencies for unigrams in tweet text, emoji, profile data, YouTube video data and image tags. Baseline {\em Model(1)} considers all 3,285 gang and non-gang member profiles in our dataset. Baseline {\em Model(2)} considers all Twitter profiles that contain every feature type discussed in Section~\ref{features}. Because a Twitter profile may not have every feature type, baseline {\em Model(1)} represents a practical scenario where not every Twitter profile contains every type of feature. However, we compare our results to both baseline models and report the improvements.

\subsection{10-fold cross validation}
We conducted 10-fold cross validation experiments to evaluate the performance of our models. We used all Twitter profiles in the dataset to conduct experiments on the five methods we used to combine word embedding vectors. For each of the five vector combination methods (as mentioned in Section~\ref{sec:approach}), we trained classifiers using each learning algorithm we considered. In each fold, the training set was used to generate the word vectors, which were then used to compute features for both the training set and test set. For each 10-fold cross validation experiment, we report three evaluation metrics for the `gang' (positive) and `non-gang' (negative) classes, namely, the Precision = \(tp / (tp + fp)\), Recall = \(tp / (tp + fn)\), and $F1$-score = \(2 * (Precision * Recall) / (Precision + Recall) \),  where \(tp\) is the number of true positives, \(fp\) is the number of false positives, \(tn\) is the number of true negatives, and \(fn\) is the number of false negatives. We report these metrics for the `gang' and `non-gang' classes separately because of the class imbalance in the dataset.

\begin{table*}[]

\begin{tabular}{|l|l|c|c|c|c|c|c|}
\hline
\multicolumn{1}{|c|}{\multirow{2}{*}{\textbf{Model}}} & \multicolumn{1}{c|}{\multirow{2}{*}{\textbf{Classifier}}} &
\multicolumn{3}{c|}{\textbf{Gang}}     & \multicolumn{3}{c|}{\textbf{Non-Gang}} \\ \cline{3-8} 
                                            &                             & \textbf{Precision} & \textbf{Recall} & \textbf{$F1$-score} & \textbf{Precision}  & \textbf{Recall} & \textbf{$F1$-score} \\ \hline
Baseline {\em Model(1)}                 & RF          & 0.8792          & 0.6374       & 0.7364         & 0.9507           & \textbf{0.9881}  & 0.9690    \\ \hline      
Baseline {\em Model(2)}                 & RF          & \textbf{0.8961}          & 0.6994       & 0.7755         & 0.9575           & 0.9873  & 0.9720    \\ \hline                                           
\multirow{3}{*}{$V_{p_{sum}}$
}                           & LR                         & 0.6007          & 0.7045       & 0.6459         & 0.9576           & 0.9346       & 0.9458        \\ \cline{2-8} 
                                            & RF                          & 0.7412          & 0.7085       & 0.7213          & 0.9596           & 0.9659       & 0.9626         \\ \cline{2-8} 
                                            & SVM                         & 0.5929          & \textbf{0.7728}        & 0.6559         & \textbf{0.9661}           & 0.9116       & 0.9369        \\ \hline
\multirow{3}{*}{$V_{p_{avg}}$ 
}                           & LR                         & 0.8394          & 0.5789        & 0.6824         & 0.9442           & 0.9850        & 0.9641        \\ \cline{2-8}    
                                            & RF                          &   0.7627         & 0.7439       & 0.7501         & 0.9650           & 0.9675       & 0.9662        \\ \cline{2-8} 
                                            & SVM                         &   0.8405        & 0.7217       &  0.7740        & 0.9624           & 0.9807       & 0.9715        \\ \hline             
\multirow{3}{*}{$V_{p_{sum(count)}}$ 
}                           & LR                         & 0.6768          & 0.6699       & 0.6681         & 0.9537           & 0.9540       & 0.9537         \\ \cline{2-8} 
                                            & RF                          & 0.7484          &    0.7346    & 0.7386         & 0.9631           & 0.9648       & 0.9639        \\ \cline{2-8} 
                                            & SVM                         & 0.5656          & 0.7180       & 0.6267          & 0.9594           & 0.9212       & 0.9395         \\ \hline
\multirow{3}{*}{$V_{p_{sum(tf-idf)}}$ 
}                           & LR                         & 0.7901         & 0.7078        & 0.7438         & 0.9595            & 0.9742       & 0.9667        \\ \cline{2-8} 
                                            & RF                          & 0.7979          & 0.7074       & 0.7470         & 0.9598           & 0.9746       & 0.9671        \\ \cline{2-8} 
                                            & SVM                         & 0.7352          & 0.6810       & 0.6952          & 0.9557           & 0.9628       & 0.9589        \\ \hline
\multirow{3}{*}{$V_{p_{avg(sum(count))}}$ 
}                      
  & LR                         & 0.8490          & 0.7327       & \textbf{0.7835}          & 0.9634            &  0.9815        & \textbf{0.9723}        \\ \cline{2-8} 
                                            & RF                          & 0.7657          & 0.7443   & 0.7519         & 0.9650           & 0.9678       & 0.9663        \\ \cline{2-8} 
                                            & SVM                         & 0.7921          & 0.7194       & 0.7500          & 0.9615           & 0.9735       & 0.9674       \\ \hline

\end{tabular}
\centering
\caption{Classification Results Based on 10-Fold Cross Validation.}
\label{results}
\vspace{1mm}
\end{table*}

\subsection{Experimental results}
Table~\ref{results} presents 10-fold cross validation results for the baseline models (first and second rows) and our word embeddings-based models (from third row to seventh row). As mentioned earlier both baseline models use a random forest classifier trained on term frequencies of unigram features extracted from all feature types. The two baseline models only differs on the training data filtering method used, which is based on the availability of features in the training dataset as described in~\cite{balasuriya2016twittergang}. The baseline {\em Model(1)} uses all profiles in the dataset and has a $F1$-score of 0.7364 for `gang' class and 0.9690 for `non-gang' class. The baseline {\em Model(2)} which only uses profiles that contain each and every feature type has a $F1$-score of 0.7755 for `gang' class and $F1$-score of 0.9720 for `non-gang' class.

Vector sum is one of the basic operations we can perform on word embedding vectors. The random forest classifier performs the best among vector sum-based classifiers where logistic regression and SVM classifiers also perform comparatively well ($V_{p_{sum}}$). Using vector mean ($V_{p_{avg}}$) improves all classifier results and SVM classifier trained on the mean of word embeddings achieves very close results to the baseline {\em Model(2)}. Multiplying vector sum with corresponding word counts for each word in word embeddings degrades the classifier accuracy for correctly identifying the positive class ($V_{p_{sum(count)}}$). When we multiply words by their corresponding $tf$-$idf$ values before taking the vector sum, we again observe an increase in the classifiers' accuracy ($V_{p_{sum(tf-idf)}}$). We achieve the best performance by averaging the vector sum weighted by term frequency ($V_{p_{avg(sum(count))}}$). Here we multiply the mean of the word embeddings by count of each word, which beats all other word embeddings-based models and the two baselines. In this setting, logistic regression classifier trained on word embeddings performs the best with a $F1$-score of 0.7835. This is a 6.39\% improvement in performance when compared to the baseline {\em Model(1)} and a 1.03\% improvement in performance when compared to baseline {\em Model(2)}. Overall, out of the five vector operations that we used to train machine learning classifiers, four gave us classifier models that beat baseline {\em Model(1)} and two vector based operations gave us classifier models that either achieved very similar results to baseline {\em Model(2)} or beat it. This evaluation demonstrates the promise of using pre-trained word embeddings to boost the accuracy of supervised learning algorithms for Twitter gang member profile classification.

\section{Conclusion and Future Work} \label{sec:con} 
This paper presented a word embeddings-based approach to address the problem of automatically identifying gang member profiles on Twitter. Using a Twitter user dataset that consist of 400 gang member and 2,865 non gang member profiles, we trained word embedding models based on users' tweets, profile descriptions, emoji, images, and videos shared on Twitter (textual features extracted from images, and videos). We then use the pre-trained word embedding models to train supervised machine learning classifiers, which showed superior performance when compared to the state-of-the-art baseline models reported in the literature. We plan to further extend our work by building our own image classification system specifically designed to identify images commonly shared by gang members such as guns, gang hand signs, stacks of cash and drugs. We would also like to experiment with automatically building dictionaries that contain gang names and gang-related slang using crowd-sourced gang-related knowledge-bases such as HipWiki\footnote{http://www.hipwiki.com/Hip+Hop+Wiki}. We also want to experiment with using such knowledge-bases to train word embeddings to understand whether having access to gang-related knowledge could boost the performance of our models. Finally, we would like to study how we can further use social networks of known gang members to identify new gang member profiles on Twitter.

\subsubsection*{Acknowledgements}
We are grateful to Sujan Perera and Monireh Ebrahimi for thought-provoking discussions on the topic. We acknowledge partial support from the National Science Foundation (NSF) award: CNS-1513721: ``Context-Aware Harassment Detection on Social Media'' and National Institutes of Health (NIH) award: MH105384-01A1: ``Modeling Social Behavior for Healthcare Utilization in Depression''. Any opinions, findings, and conclusions/recommendations expressed in this material are those of the author(s) and do not necessarily reflect the views of the NSF or NIH.

\bibliographystyle{alpha}
\input{paper2p.bbl}


\end{document}

%% file: paper2p.bbl
\newcommand{\etalchar}[1]{$^{#1}$}